\documentclass[preprint,aps,prl,showpacs,superscriptaddress]{revtex4}
\usepackage{amssymb}

\usepackage{graphicx}

\begin{document}
\title{Secure and efficient decoy-state quantum key distribution with inexact pulse intensities}
\author{Xiang-Bin Wang}\email{xbwang@mail.tsinghua.edu.cn}
\affiliation{Department of Physics, Tsinghua University, Beijing
100084, China}

\begin{abstract}
 We present a
general theorem for the efficient verification of the lower bound of
single-photon transmittance. We show how to do decoy-state  quantum
key distribution efficiently with large random errors in the
intensity control. In our protocol, the linear terms of fluctuation
disappear and only the quadratic terms take effect. We then show the
unconditional security of decoy-state method with whatever error
pattern in intensities of decoy pulses and signal pulses provided
that the intensity of each decoy pulse is less than $\mu$ and the
intensity of each signal pulse is larger than $\mu'$.
\end{abstract}


\pacs{
03.67.Dd,
42.81.Gs,
03.67.Hk
}
\maketitle


{\em Introduction.---\/} The decoy-state method\cite{H03, Wang05,
Wang05_2,LMC05,HQph} or some other methods\cite{scran,kko,zei} can
be used for two remote parties, Aice and Bob  to do secure quantum
key distribution (QKD)\cite{BB84,GRTZ02,DLH06}  Even Alice only uses
an imperfect source, e.g., a coherent light\cite{QKD,PNS,PNS1}.
A secure final key can be distilled by using the separate
theoretical results\cite{GLLP04} if one knows the upper bound of the
fraction of tagged bits (those raw bits generated by multi-photon
pulses from Alice) or equivalently, the lower bound of the fraction
of un-tagged bits (those raw bits generated by single-photon pulses
from Alice). The goal of decoy-state method is to verify such bounds
faithfully and efficiently. \\Recently, a number of experiments on
decoy-state QKD have been done\cite{Lo06,peng,ron}. However, the
existing theory of decoy-state method assumes the exact control of
pulse intensities. A new problem arose in practice is how to carry
out the decoy-state method efficiently given the inexact control of
pulse intensity.
In this Letter, we study this problem and we find that if the
intensity of each pulses are bounded in a reasonable range, we can
still verify the fraction of single-photon counts efficiently.
\\{\em General idea.---\/}
There are two goals here, security and efficiency. For security, the
verified value of fraction of single-photon counts from our method
must never larger than the true value given whatever channel. For
this part we should not assume any specific property for the
channel. This section will give  a general method for secure
verification of fraction of un-tagged bits. We also want our
protocol to be efficient. We want that, in the normal situation
where there is no Eve, the verified value of the fraction of
single-photon counts is rather close to the true value.
We shall  evaluate the efficiency of our protocol in another
section. \\We start from the definition of the {\em counting rate}
of certain pulses. Given a class of $N$ independent pulses, after
Alice transmits them to Bob one by one, if Bob observes $n$ counts
at his side, the counting rate for pulses in this class is $s=n/N$.
If the state of source in photon-number space is known, the fraction
of single-photon counts is known given the counting rate of all
those single-photon pulses. We shall only consider how to find the
single-photon pulse counting rate hereafter.
 Suppose there
are $l$ different subclasses of independent light pulses in a
certain class. We denote the fractions of pulses in each subclasses
by $a_0,a_1\cdots,a_l$. If all these pulses are sent to Bob through
whatever channel, the total counts observed by Bob should be equal
to the summation of the counts due to the pulses of each subclasses.
 Therefore we have
\begin{equation}\label{ele}
S=\sum_0^l a_is_i
\end{equation}
$S$ is the counting rate of the whole class while $s_i$ is the
counting rate of the $i$th class.

The decoy-state method itself does {\em not} require the Possonian
distribution of source light, though it has been applied to the case
of Possonian distribution\cite{H03, Wang05, Wang05_2,LMC05,HQph}.
Most generally, in a 3-intensity decoy-state protocol, we consider 3
classes of states, $Y_0,Y,Y'$. $Y_0$ contains all vacuum pulses. $Y$
contains three subclasses $y_0,y_1,y_c$ for vacuum pulses,
single-photon pulses and multi-photon pulses, respectively. Classes
$Y'$ contains $4$ subclasses, $y_0',y_1',y_c',y_d'$. We shall use
notations $S_0,S,S'$ for counting rates of classes of $Y_0,Y,Y'$,
respectively; notations $\{s_x\},\{s'_x\}$ for counting rates of
subclasses $\{y_x\},\{y'_x\}$ and $x$ can be {\em 0,1,c,d}.  Using
eq.(\ref{ele}) we have
\begin{eqnarray} \label{ele1}
\begin{array}{ll}
S= a_0s_0+a_1s_1+a_cs_c\\
S'=a_0's_0'+a_1's_1'+a_c's_c'+a_d's_d' \end{array} .\end{eqnarray}
We shall regard  $S_0,S,S'$ as {\em known} parameters since they are
observed directly in the protocol.  In general, $s_x\not= s_x'$.
Since all of them are non-negative, we can assume
\begin{equation}\label{ele2}
s_1'=(1-r_1)s_1, ~~ s_c'=\omega_c s_c
\end{equation}
and $(1-r_1),\omega_c$ are non-negative numbers. If we define
$b_c'=\omega_ca_c'$, eqs.(\ref{ele1}) is equivalent to
\begin{eqnarray}\label{num1}
\left\{\begin{array}{ll}
E=a_1s_{1}+a_cs_{c} \\
E'= a_1's_1 + b_c's_c
\end{array} \right.
\end{eqnarray}
and $E=S-a_0s_0$; $E'=S'-b'_0s'_0+f_1-a_d's_d'$ and
$f_1=r_1a_1's_1$. Therefore, it will be secure if we find the
smallest value $s_1$ satisfying the equation above among all
possible values for parameters $E,E', a_1,a_c,a_1',b_c'$. In
general, this can be done numerically.
  To seek the lower bound of $s_1$ based on eqs.(\ref{num1}),
  we need first find the ranges of all parameters. As we are going
  to show, the parameters of $\{a_x,a'_x\}$ can be determined rather precisely by
  a type of tomography.
 In our protocol, we mix all pulses from 3 classes randomly and we can simply deduce
 $\omega_{c},f_1$ by classical  random
sampling theory. In a decoy-state method, we let subclass
$y_0(y_0'),y_1(y_1')$ contains all those vacuum pulses,
single-photon pulses from class $Y(Y')$, $y_c$ contains all those
multi-photon photon pulses from $Y$. Suppose the state of
multi-photon pulses from $Y$ are $\rho_c$ and state of class $Y'$ is
a convex form of $\rho_c$ and other states. We require the state of
a pulse from $y_c'$ be also $\rho_c$, same to that of $y_c$. We
emphasize that in the protocol Alice does not need to know which
pulse belongs to which subclass, we only need that mathematically
there exists such subclasses\cite{Wang05}.

\begin{figure}
\centerline{\includegraphics[scale=0.7]{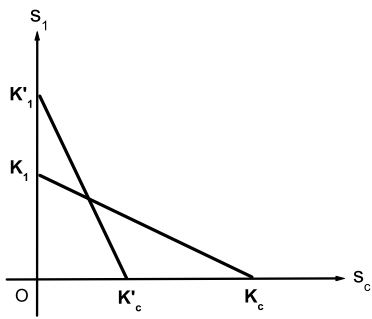}}
\caption{\label{Fig1} Graphics of eqs.(\ref{num1}) in $s_1-s_c$
plane. Obviously, $s_1$ value will be raised if $K_1$ or $K_c$ is
raised, or if $K_1'$ or $K_c'$ is decreased. This leads to our
theorem 1.}
\end{figure}

\begin{figure}
\centerline{\includegraphics[scale=0.60]{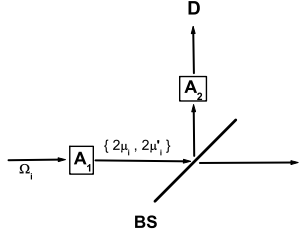}}
\caption{\label{Fig2} Our proposed set-up for decoy-state QKD. D:
detector, A: attenuator, BS: 50:50 beam-splitter. The transmitted
light is sent to Bob and the reflected light is detected by
Alice.}
\end{figure}

 For certain two subclasses, if each pulses are
independent and the states for pulses of two subclasses are same,
the pulses of one class can be regarded as samples of all pulses of
both classes, if all pulses are randomly mixed. Therefore, if each
pulses of classes $Y,Y'$ are independent and randomly mixed, the
counting rates for pulses of subclasses $\{y_1,y_1'\}$,
$\{y_c,y_c'\}$, $\{y_0,Y_0\}$ and $\{y_0',Y_0\}$ can only be
different by a statistical fluctuation.
Therefore, bounds of $s_0,s_0'$ are known and parameters of
$r_1,\omega_c$ can be formulated by $s_1,s_c$ and the number of
pulses from classical sampling theory\cite{Wang05}. If there are a
larger number of pulses, counting rates of the same state from
different classes should be almost the same.
For the case of using exact intensities of $0,\mu,\mu'$, the
parameters of $\{a_x\}$ and $\{a_x'\}$ are known from the
information of the source state. For example, given coherent light
of intensity $0,\mu,\mu'$ for classes $Y_0,Y,Y'$, respectively, we
have\cite{Wang05}
\begin{eqnarray}\label{ip}\begin{array}{llll}
a_0=A_0= e^{- \mu}; a_1=A_1= \mu e^{-\mu},\\
a_c=A_c=1-e^{- \mu}-\mu e^{-\mu}\\a_0'=A_0'= e^{-
\mu'}, a_1'=A_1'= \mu' e^{- {\mu}'}\\
{b_c}'=\omega_c{A_c}' =\omega_c\frac{ \mu'^2 e^{- {\mu}'}} {{ \mu}^2
 e^{-\mu}}A_c
\end{array} \label{ideal}\end{eqnarray}
\\
 {\em A theorem for calculation of $s_1$.---\/}
 Most directly, given the ranges of each parameters
 involved in our protocol, we can solve Eqs.(\ref{num1})
 numerically for the lower bound of single-photon counts. However, since here
 there are a number of parameters, the numerical complexity can be huge.
 We can avoid the complexity by the following treatment.
 Define $K_1=\frac{E}{a_1}$, $K_c=\frac{E}{a_c}$,
$K_1'=\frac{E'}{a_1'}$, $K_c'=\frac{E'}{b_c'}$. We can always find a
meaningful solution for $s_1,s_c$ if   \begin{equation}K_1'>
K_1>0,~~ K_c>K_c'>0.\label{condition}\end{equation}  As it is shown
in Fig.(\ref{Fig1}), the solution of $s_1,s_c$ is the crossing point
of the two lines in $s_c-s_1$ plane. In this plane, it is easy to
see that $s_1$ value rises if $K_1'$ or $K_c'$ decreases, or if
$K_1$ or $K_c$ rises.
Therefore, the largest possible values of $K_1',K_c'$ and the
smallest possible values of $K_1,K_c$ will produce the lower bound
of $s_1$.  We have the following  theorem: {\bf Theorem 1:} Given
eqs.(\ref{num1}), if eqs.(\ref{condition}) holds,
 the maximum of
values of $a_0s_0, a_1,a_c,f_1$ and minimum values of
$a_0's_0',a_1',b_c',a_d'$ will give the smallest  result of $s_1$ in
eqs.(\ref{num1}). An alternative proof is shown in the appendix.
\\{\em Decoy-state QKD with simple tomography.---\/}
We assume that the intensity fluctuation of each individual pulse is
random. Consider a protocol where Alice controls the intensity by a
feedback circuit. Each time she first produces a father pulse $F_i$
whose intensity is not known exactly. This pulse is then split into
two daughter pulses: $D_{i}$ and $\Omega_{i}$. The intensity of
pulse $D_{i}$ is detected (e.g., by homodyne measurement) and this
detection outcome determines the instantaneous attenuation to
$\Omega_{i}$ to obtain the supposed intensity. There could be {\em
random} errors in detecting $D_{i}$, in instantaneously controlling
the attenuator. (The feedback circuit is not drawn in
Fig.(\ref{Fig2}) ).

 Whenever Alice wants to use $\mu$ or $\mu'$, she actually uses
\begin{equation}
\mu_i=(1+\delta_i)\bar \mu; ~~ \mu_i'=(1+\delta_i')\bar \mu' .
\label{cons}\end{equation} She does not know each specific value
of $\delta_i$ or $\delta_i'$. But as we shall show she can know
the averaged value of
\begin{equation}
\bar \mu = \frac{1}{N}\sum_1^{N} \mu_i;~~ \bar \mu' =
\frac{1}{N}\sum_1^{N'} \mu'_i
\end{equation}
rather exactly. Here $N,N'$ are number pulses in class $Y,Y'$,
respectively.
   Moreover, given the fact
\begin{equation}\label{zero}
\sum_0^N\delta_i=\sum_0^{N'}\delta'_i=0
\end{equation}
Alice can find rather narrow ranges
 for relevant parameters of her states by a type of simple tomography.
 She can, as shown in Fig.(\ref{Fig2}) , every time first produces a pulse of
intensity $2\mu_i$ or $2\mu_i'$ by attenuation. The pulse is then
split by a 50:50 beam-splitter. The transmitted mode is sent to Bob,
the reflected mode goes to a low efficient photon detector, e.g., a
detection efficiency of $\xi\le  10\%$.
  We shall simply use the mathematical model of an attenuator
with transmittance $\xi$ and a perfect yes/no detector.
Suppose she has observed the clicking rate of $h+d_0$ and $h'+d_0$
for those $N$ reflected pulses of intensity $\{\mu_i\}$ and $N'$
reflected pulses of intensity $\{\mu_i'\}$, respectively. Here $d_0$
is the dark count rate of her detector. Mathematically,
\begin{equation}
\sum_0^{N}(1-e^{-\xi\mu_i})/N = h\label{Taylor}
\end{equation}
This leads to the following facts after Taylor expansions:
\begin{equation}\label{ine0}
\bar\mu \ge h/\xi;\end{equation}
\begin{equation}\label{ine1}
\bar\mu\le \mu_+=\frac{1-\sqrt{1-2h(1+\zeta)}}{\xi(1+\zeta)}\approx
h/\xi+h^2(1+\zeta)/(2\xi)\end{equation} and notation
$\zeta=\sum\delta_i^2/N \le \delta^2, \delta =Max\{|\delta_i|\}$.
Combine Eqs(\ref{ine0},\ref{ine1}) and the Taylor expansion of
Eq.(\ref{Taylor}) we obtain an even more tightened lower bound
formula
\begin{equation}\label{ine00}
\bar\mu\ge \mu_-=h/\xi+h^2/(2\xi)-\xi^2\mu_+^3/3!
\end{equation}
Replacing
$h$ with $h'$ in Eqs.(\ref{ine1},\ref{ine00}) we can also bound
$\bar \mu'$ by $\mu_-'\le \bar\mu'\le\mu_+'$.
Similarly, we shall use $\zeta'=\sum {\delta_i'}^2/N'\le
{\delta'}^2,~~\delta'=Max\{\delta_i'\}$.
Later,  she can verify the bounds of all parameters with the
observed values $h,h'$ and the above formulas for
$\bar\mu,\bar\mu'$. The true state for a pulse in class $Y$ is
\begin{equation}
 \frac{1}{N}\sum_{i,n=0}^{N,\infty}
\frac{\mu_i^n e^{-\mu_i}}{n!}|n\rangle\langle n|=a_0
|0\rangle\langle 0| + a_1|1\rangle\langle 1| + a_c  \rho_c
\end{equation}
and $a_0,a_1,$ are $\sum_i {e^{-\mu_i}}/N, \sum_i
{\mu_ie^{-\mu_i}}/N, a_c=1-a_0-a_1$. Here $\rho_c $ is the
averaged state of all multi-photon pulses in class $Y$. Obviously,
if $\bar\mu'$ is sufficiently large than $\bar\mu$ and the
intensity error is not {\em too} large, we can also write
$\rho_{\mu'}$ in a convex form including $\rho_c$:
\begin{equation}\rho_{\mu'}= a_0' |0\rangle\langle 0| + a_1'|1\rangle\langle 1|
+ a_c'  \rho_c + a_d' \rho_d
\end{equation}
and $a_0',a_1'$ are $ \sum_i e^{-\mu_i'}/N', \sum_i
\mu_i'e^{-\mu_i'}/N'$,
  $a_c'=\frac{\sum\mu_i'^2e^{-\mu_i'}/N'}{\sum\mu_i^2
e^{-\mu_i}/N}a_c$, $a_d'\ge 0$, $\rho_d$ {\em is} a density
operator.
We have the following bound values those parameters involved
\begin{eqnarray}\label{pbound0}
\left\{\begin{array}{lll} e^{-\mu_+}\le a_0 \le e^{-\mu_-}
(1+\bar\mu^2\delta^2/2)
\\ (1-\mu_-\delta^2)\mu_-e^{-\mu_-}\le a_1
\le \mu_+e^{-\mu_+}
\\ a_c
\le 1-e^{-\mu_+}-\mu_+e^{\mu_+}+\mu_+\delta^2
\end{array} \right.
\end{eqnarray}
\begin{eqnarray}\label{pbound1}
\left\{\begin{array}{llll} a_0' =\frac{1}{N}\sum e^{-\mu_i'}\ge
e^{-\bar\mu'_+}
\\ a_1'
\ge (1-\mu_-'\delta'^2)\mu_-'e^{-\mu_-'}
\\ b_c'
\ge \omega_c\frac{\mu_-'^2[1-e^{-\mu_-}-\mu_-e^{-\mu_-}]}{(1+\delta^2)\mu_+^2e^{\mu_-'-\mu_+}}\\
a_d'\ge 0
\end{array} \right.
\end{eqnarray}
{\em Efficiency evaluation.---\/}
 We shall compare the efficiencies of two
protocols, the ideal protocol where the intensity of every light
pulse in class $Y$ or $Y'$ is exactly $\mu$ or $\mu'$ and our
protocol where the intensity of each light pulses is inexactly
controlled. In a real experiment using our protocol, Alice simply
reads $h,h'$ values and then calculate the lower bound of $s_1$.
Here  we assume the model that Alice has observed
\begin{equation}\label{model}
h=\xi\mu-\xi^2\mu^2/2;~~h'=\xi\mu'-\xi^2\mu'^2/2.
\end{equation}
Given these, we can calculate bounds for $\bar\mu,~~\bar\mu'$ by
our earlier equations.  We take the following assumptions:
$\mu=0.2$, $\mu'=0.6$, $\xi=5\%$ for Alice's detection efficiency,
linear channel with transmittance $\eta=10^{-4}, S_0=s_0=s_0'=0, ~
N=10^9$ and $\delta=\delta'$. In both protocols we use $f_1\le
10a_1\sqrt{\frac{s_1}{N\mu e^{-\mu}}}$ and $\omega_c \ge
1-10\sqrt{\frac{1}{s_c(1-a_0-a_1)N}} $.
To compare the efficiencies of our protocol and the ideal
protocol, we only need to compare solutions of eqs.(\ref{num1})
for two protocols.
We now denote $s_1,\tilde s_1$ to be the results of single-photon
transmittance from our protocol and the ideal protocol,
respectively.  The fraction of un-tagged bits from class $Y'$ is
given by
\begin{eqnarray}\left\{\begin{array}{ll}
\Delta_1'=s_1A_1'(1-\mu\delta^2)/(1-e^{-\eta\mu'})\\
\tilde\Delta_1'=\tilde s_1 A_1'/(1-e^{-\eta\mu'})
\end{array}\right.\end{eqnarray}
 $\Delta_1'$ is
for our protocol, $\tilde \Delta_1'$ is for the ideal protocol.
We shall calculate $T=s_1/\tilde s_1,
~R=\Delta_1'/\tilde\Delta_1'$. We find very good results given
various $\delta$ values. (See details in table 1.)
\begin{table}
\caption{\label{tab:table1} Efficiency comparison of our protocol
and an ideal protocol.
}
\begin{ruledtabular}
\begin{tabular}{llllllll}
$\delta$ &5\% & 10\% & 15\%& 20\% & 25\%&30\% & 35\% \\
\hline $T$& 99.8\%&99.6\%&99.2\%&98.7\%& 98.0\%&97.2\% &96.3\%\\
\hline
$R$ &99.7\%& 99.0\% &97.9\% & 96.3\%&94.4\% &91.9\%& 89.2\%\\
\end{tabular}
\end{ruledtabular}
\end{table}
Moreover, the results our protocol can be even improved because
there are obviously better ways to bound $\zeta,\zeta'$ more
tightly. For example, suppose we know that the fluctuation of more
than $90\%$ of the pulses is less than $10\%$, even though the
largest fluctuation is $50\%$, we have $\zeta\le 3.4\%$ and we can
verify a $R\ge 96\%$ with  $\delta^2$  being replaced by $\zeta$
in all equations. For another example, Alice can use two detectors
of efficiency $\xi_1,\xi_2$ to tightly verify the upper bound of
$\zeta$: Every time she first produces a pulse of intensity $3x$,
($x$ can be 0, around $\mu$ or $\mu'$). She equally divides the
pulse into 3 modes, mode b ia sent to Bob, modes 1 and 2 are sent
to detector 1 and 2 respectively. Using the number of counts of
each detector, she can verify an upper bound of $\zeta$ value only
{\em a little bit} larger than the true value of $\zeta$. (This
will be reported elsewhere separately.)

 Our
theorem 1 is based on the conditions of eqs.(\ref{condition}). These
conditions are related to the statistical fluctuations which are
dependent on the value of $s_1,s_c$. But we can verify these
conditions {\em before} knowing the exact values of $s_1,s_c$.
First, we assume $s_c> 2\eta $. This assumption leads to $s_1<\eta
$. Here $\eta$ is the channel transmittance. We can assume so
safely.  If the assumption $s_c> 2\eta $ is incorrect, then
$s_1>\eta$ which is a quite good result.
If the assumption of $s_c>2\eta$ is correct, then our calculation
based on this is alright. In whatever case, it is secure if we use
the assumption for calculation and we then use $Min\{\eta,s_1\}$
($s_1$ is the calculated result.) Therefore we can have bound values
of
\begin{equation}
f_1\le 10a_1\sqrt{\frac{\eta}{a_1N}};~~\omega_c \ge
1-\sqrt{\frac{1}{2\eta a_c N}}.
\end{equation}
Given these, we can easily verify eqs.(\ref{condition}) and then use
our
 theorem 1 safely.
\\{\em Effect of inexact vacuum pulses in class $Y_0$.---\/}
In general, $S_0\not=0$. We can safely set $s_0'=0$ according to our
theorem 1 and  we only need to consider the upper bound of $s_0$.
Asymptotically,  we can simply replace $s_0$ by $S_0$ even though
pulses in $Y_0$ are not strictly vacuum.
Let's assume the actual state in $Y_0$ is
$\rho_0=(1-\epsilon_0)|0\rangle\langle 0| + \epsilon_1
|1\rangle\langle 1| + \epsilon_m \rho_m$. Here $\rho_m$ is a state
of multi-photon pulses, $\epsilon_m = O(\epsilon_1^2),
\epsilon_1<<1$ and $\epsilon_0= \epsilon_1+\epsilon_m$. Therefore,
we have
\begin{equation}\label{dark}
S_0 =(1-\epsilon_0)s_0 +\epsilon_1s_1+\epsilon_m s_m.
\end{equation}
This leads to a preliminary upper bound of $ s_0\le
\frac{S_0}{1-\epsilon_0}. $ We then replace $s_0$ in
eqs.(\ref{num1}) and solve the equation for lower bound of $s_1$. We
assume $s_1\ge 1.5S_0$ at this stage, otherwise the protocol should
be discarded.
 Now we
consider eq.(\ref{dark}) again. We have a new bound of $ s_0\le
\frac{S_0}{1-\epsilon_0} -\epsilon_1s_1 \le S_0. $
\\{\em The unconditional security for whatever error pattern.---\/} Suppose we don't use
the feedback control for $\Omega_i$ in Fig.(\ref{Fig2}). Most
generally, the intensity fluctuation of each pulses is not perfectly
random.
Now the probability for a pulse from $y_x$ or from $y_x'$ can
change slightly at different time intervals
therefore $s_x$ can be slightly different from $s_x'$ in the whole
time series even there is no statistical fluctuation. For example,
it is possible that in a certain time interval,  the probability of
using $y_1$ $(y_1')$ is less (larger) than the averaged probability
of using $y_1$ $(y_1')$, Eve can produce a certain time-dependent
channel transmittance for those single-photon pulses sent from Alice
and the averaged counting rates of $y_1$ and $y_1'$ in the whole
time series can be different from each other, even there is no
statistical fluctuation. This is to say, in general, pulses of
sub-class $y_x$ and $y_x'$ in principle can {\em not} be regarded as
{\em randomly mixed} if the intensities of each pulses are not
exactly controlled.
We need a separate security proof for a protocol with {\em whatever}
pattern of intensity error. We now prove that the protocol is secure
if $\mu_i\le \mu$ and $\mu_i'\ge \mu'$.

We start from a virtual protocol, {\em Protocol 1:}  At each time
$i$ in sending a pulse to Bob, Alice produces a  bipartite state
\begin{equation}
\rho_i(2)=p_0 |z_0\rangle\langle z_0|\otimes|0\rangle\langle 0|+p
|z_1\rangle\langle z_1|\otimes\rho_\mu + p'  |z_2\rangle\langle
z_2|\otimes\rho_{\mu_i'} \end{equation} and announces the value of
$\mu_i'$. Here $\rho_x=\sum_{n=0}^\infty
\frac{x^ne^{-x}}{n!}|n\rangle\langle n|$, the value $\mu$ keeps to
be constant but $\mu_i'$ can change from time to time and $\mu_i'$
is not less than a constant value $\mu'$. States $\{|z_x\rangle\}$
are orthogonal to each other for different $x$ ($x=0,1,2$) and
$p_0+p+p'=1$.  Alice keeps the light pulse in the first subspace and
sends out the pulse in the second subspace of the bipartite state to
Bob, $i$ runs
 from 1 to $N_t$, the number of total pulses sent to Bob.
  Later, Alice measures her states
($\{|z_x\rangle\}$) and she can know which  pulse in the second
subspace of the bipartite state belongs to which class ($Y_0,Y$ or
$Y'$). As we have shown in Eqs.(\ref{ip}), state $\rho_\mu$ can be
written in the convex form of $\rho_\mu=A_0|0\rangle\langle 0| +A_1
|1\rangle\langle 1| +A_c \rho_c$ and $A_c\rho_c =\sum _2^\infty
\frac{\mu^n e^{-\mu}}{n!}|n\rangle\langle n|$.
Since $\mu_i'\ge \mu'$, we always have the following convex form for
state $\rho_{\mu_i'}$
\begin{equation}
\rho_{\mu_i'}=A_1' |1\rangle\langle 1| + A_c' \rho_c+
(1-A_1'-A_c')\rho_{e}^{i}
\end{equation} where $A_1'=\mu' e^{-\mu'}$, $A_c'=\frac{A_c\mu'^2
e^{-\mu'}}{\mu^2 e^{-\mu}}$.
 Obviously, the specific formula
 for $\rho_{e}^i$
 exists but it is unimportant here since we only need
 the fact that $\rho_{e}^i$  is a density operator\cite{Wang05}.
 To anybody outside Alice's lab, Alice {\em could } have used a
 tripartite state of
 \begin{eqnarray}
\rho_i(3) = p_0 |z_0\rangle\langle z_0|\otimes |z_0\rangle\langle
z_0|\otimes |0\rangle\langle 0|\nonumber\\+p|z_1\rangle\langle
z_1|\otimes \left(A_0 |v_0\rangle\langle v_0|\otimes
|0\rangle\langle 0|+A_1 |v_1\rangle\langle v_1|\otimes
|1\rangle\langle1|+A_c |v_c\rangle\langle v_c|\otimes
\rho_c\right)\nonumber
\\+p'|z_2\rangle\langle z_2|\otimes \left[A_1' |v_1'\rangle\langle
v_1'|\otimes |1\rangle\langle1|+A_c' |v_c'\rangle\langle
v_c'|\otimes \rho_c'+ (1-A_1-A_c')|v_e'\rangle\langle v_e'|\otimes
\rho_e^i\right]
 \end{eqnarray}
and  those states in the second subspace are all orthogonal to each
other. Alice keeps the pulses in the first and second subspaces and
sends out the pulse in the third subspace to Bob. Given this, we can
define 3 classes $Y_0,Y,Y'$ of pulses: if Alice obtained her
measurement outcome of $|z_0\rangle,|z_1\rangle$ or $|z_2\rangle$ in
the first subspace, the corresponding pulse sent out is regarded as
a pulse of class $Y_0,Y$ or $Y'$. We can also define 3 subclasses
$y_0,y_1,y_c$ of $Y$ and  3 sub-classes $\tilde y_1',\tilde y_c',
y_e'$ of $Y'$: if Alice obtains her measurement outcome of
$|v_0\rangle,|v_1\rangle$ or $|v_c\rangle$ in the second subspace,
the corresponding pulse sent out is regarded as a pulse of sub-class
$y_0,y_1$ or $y_c$; if Alice obtains her measurement outcome of
$|v_1'\rangle,|v_c'\rangle$ or $|v_e'\rangle$ in the second
subspace, the corresponding pulse sent out is regarded as a pulse of
sub-classes $\tilde y_1', \tilde y_c'$ or $y_e'$. Here $\tilde y_1'$
is a bit different from the sub-class $y_1'$ defined before: $y_1'$
defined before contains all those single-photon pulses of $Y'$ while
$\tilde y_1'$ here possibly does not contain all single-photon
pulses in $Y'$ if $\mu_i'< \mu'$, since some of single-photon pulses
from $Y'$ are regarded as elements of $y_e'$ now, according to our
definition. Similarly, $\tilde y_c'$ here is also a bit different
from $y_c'$ as defined before. Since pulses of sub-classes $\tilde
y_1',\tilde y_c'$ occur with {\em constant} probabilities, pulses
from sub-class $y_1,\tilde y_1'$, pulses from $y_c,\tilde y_c'$ and
pulses from $y_0,Y_0$ are {\em randomly mixed}.  For simplicity in
presentation, we only consider the asymptotic case here, i.e., the
counting rates for two sub-classes containing the same state must be
equal to each other. We can use the following constraints to verify
the single-photon transmittance $s_1$:
\begin{eqnarray}\label{num5}
\left\{\begin{array}{ll}
A_1s_{1}+A_cs_{c}=E \\
  A_1's_1 + A_c' s_c \le S'
\end{array} \right.
\end{eqnarray}
and $E=S-e^{-\mu}s_0$, $S,S'$ are the counting rates of  classes
$Y,Y'$, $s_1$ is the counting rate of class $y_1$ or $\tilde y_1'$,
$s_c$ is the counting rate of class $y_c$ or $\tilde y_c'$, $A_1=\mu
e^{-\mu}$, $A_c=1-A_0-A_1$. The value $s_0$ can be deduced from the
observed counting rate of class $Y_0$ by classical sampling theory.
In obtaining the second constraint above, we have used the fact that
$N_t p'(A_1's_1+A_c's_c ) \le N_t p' S' $, i.e., the number of
counts caused by part of pulses ($y_1'\cup y_c'$) of class $Y'$
cannot be larger than the number of counts caused by all pulses of
class $Y'$. Here, in using Eqs.(\ref{num5}), Alice actually does not
need any information of which pulse belong to which sub-class.
Therefore she can discard the pulse in the second subspace of the
tripartite state $\rho_i(3)$, consequently, she can just use the
bipartite state $\rho_i(2)$ and obtain $s_1$ value through
Eqs.(\ref{num5}).  In this protocol,  Alice announces $\mu_i'$ value
at each time but it is still secure since her announcement does not
change the fact that pulses of each sub-classes
$y_0,y_1,y_c,y_1',y_c'$ will occur with  constant probabilities
therefore classical randomly sampling theory  works, so that
Eqs.(\ref{num5}) holds. (Definitely, the protocol is also secure if
Alice does not announce $\mu_i'$ value at each time.)

Suppose in another protocol, {\em Protocol 2}, Alice uses source
state $\gamma_i$ which can in principle be obtained through
attenuating $\rho_i(2)$ in the second subspace by a factor $\chi_i$.
If Eve can attack this protocol effectively with scheme $\cal A$
then Eve can also attack {\em Protocol 1} effectively by first
attenuating the pulses
 by a time-dependent factor $\chi_i$ and then using scheme $\cal A$. Given this
fact, we conclude that any source can be used securely if that
source can in principle be obtained through attenuating state
$\rho_i(2)$ in the second subspace. This gives rise to {\bf Lemma
1:} Alice can use Eqs.(\ref{num5}) safely if the source she has
actually used in principle can be produced by attenuating
$\rho_i(2)$ in the second subspace.  This leads to {\bf Lemma 2}:
Alice can safely use Eqs.(\ref{num5}) for lower bound value of $s_1$
if she actually at each time had used any state $W_i=p_0
|z_0\rangle\langle z_0|\otimes|0\rangle 0|+p |z_1\rangle\langle
z_1|\otimes\rho_{{ \nu_i}} + p' |z_2\rangle\langle
z_2|\otimes\rho_{{ \nu_i'}}$ provided that $\nu_i\le \mu$ and
$\nu_i'\ge \mu'$. Proof: We denote the (time-dependent) attenuation
factor $\omega_i=\frac{\nu_i}{\mu} $. In protocol 1, we can set
$\mu_i' =\frac{\nu_i' \mu}{\nu_i}$ for the bipartite state
$\rho_i(2)$ and the protocol with such a setting is secure since
$\frac{\nu_i' \mu}{\nu_i}\ge \nu_i'\ge\mu'$. After attenuating
$\rho_i(2)$ by the factor $\omega_i$ in the second subspace,
$\rho_i(2)$ is changed to state $W_i$. According to our lemma 1,
Alice can use $W_i$ directly and uses Eqs.(\ref{num5}) for lower
bound of $s_1$. Moreover, it is of no difference if Alice measures
her states $\{|z_x\rangle\}$ in the very beginning. If she does
this, the protocol with source state $W_i$ is changed into a
3-intensity protocol with intensities $0,\{\nu_i\}, \{\nu_i'\}$ and
$\nu_i\le \mu$, $\nu_i'\ge \mu'$, with probability $p_0,p,p'$ for
using each of them at each time. Consequently we arrive at  {\bf
Theorem 2:} The 3-intensity protocol is secure with whatever error
pattern for intensities of decoy pulses (class $Y$) and signal
pulses (class $Y'$) provided that 1) the intensity of each decoy
pulses is less than $\mu$ and the intensity of each signal pulses is
larger than $\mu'$; 2) we use Eqs.(\ref{num5}) to calculate $s_1$.
Our result here can obviously be extended to the non-asymptotic
case. To do so, we only need to 1) replace $s_1,s_c$ by $s_1',s_c'$
in the second constraint of Eqs.(\ref{num5}); 2) give the  possible
ranges for difference between $s_1$ and $s_1'$ and difference
between $s_c,s_c'$ with exponential certainty by classical random
sampling theory\cite{Wang05}; 3) solve Eqs.(\ref{num5}) numerically
in the ranges and find the smallest $s_1$.

Although the method shown above is unconditionally secure, in the
efficiency criterion, we can have a better choice, e.g., we use the
protocol presented in Ref\cite{137}. However, there we request using
the same father pulse and exact control of attenuation. Here in
Eqs.{\ref{num5}} we don't need these and it is unconditionally
secure. The result here can apply to all existing experiments
immediately, i.e., we only need to redo the calculation of $s_1$
using our method and the existing experimental data but we don't
have to redo the experiment itself.

In summary, we have shown that decoy-state method QKD is secure and
efficient even there are errors in the intensity control.

\section{Appendix I}
Suppose $S_1,S_c$ are solution of Eq.(\ref{num1}). Then $S_1$ must
satisfy
\begin{eqnarray}
S_1 = \frac{b_c'E-a_cE'}{a_1b_c'-a_1'a_c}  \label{gen}
\end{eqnarray}
 Consider another
set of parameters $\{\tilde a_x\le a_x ,\tilde a_x'\ge a_x'\},\tilde
b_c'\ge b_c', \tilde f_1\le f_1, \tilde s_0 \le s_0, \tilde s_0' \ge
s_0'$. We define $\tilde E' = S'-\tilde a_0'\tilde s_0'+\tilde
f_1\le E'$, $\tilde E = S-\tilde a_0\tilde s_0\ge E$. We suppose
$\tilde s_1,\tilde s_c$ are solution for eqs.(\ref{num1}) with those
tilde parameters. Therefore $\tilde s_1$ should satisfy
\begin{equation}
\tilde s_1 = \frac{\tilde b_c'\tilde E-\tilde a_c\tilde E'}{\tilde
a_1\tilde b_c'-\tilde a_1'\tilde a_c} \ge \frac{\tilde b_c' E-\tilde
a_c E'}{ a_1\tilde b_c'- a_1'\tilde a_c}
\end{equation}
Since $\tilde b_c'\ge b_c',\tilde a_c \le a_c$, we can assume
$\tilde b_c' =(1+\lambda_1 )b_c$, $\tilde a_c =(1+\lambda_2 )a_c$
and $\lambda_1,\lambda_2 \ge 0$. Also we denote
$\chi=\frac{b_c'E}{a_cE'}, \gamma = \frac{a_1b_c'}{a_ca_1'}$ and we
have
\begin{eqnarray}\begin{array}{ll} \tilde s_1\ge S_1 (1+\frac{\lambda_1\chi+\lambda_2}{\chi-1}-
\frac{\lambda_1\gamma+\lambda_2}{\gamma-1})
\\
=S_1(1+\frac{(\lambda_2+\lambda_1)(\gamma-\chi)}{(\chi-1)(\gamma-1)}).\label{second}
\end{array}\end{eqnarray} While we know that
$\frac{\gamma}{\chi}=\frac{a_1b_c'/(a_1'a_c)}{b_c'E/(a_cE')}=\frac{K_1'}{K_1}>1$,
eqs.(\ref{second}) is changed to $ \tilde s_1\ge S_1. $ This
completes the proof of our theorem.
\section{appendix II}
In this appendix we derive the inequalities of
(\ref{pbound0},\ref{pbound1}). First, $a_0=\frac{1}{N}\sum
e^{-\mu_i}=\frac{1}{N}e^{-\bar\mu}\sum e^{-\bar\mu\delta_i}$.
After the Taylor expansion, we have
\begin{equation}
\sum e^{-\bar\mu\delta_i}=\sum(1-\bar\mu\delta_i
+\frac{\bar\mu^2\delta_i^2}{2}-\cdots).
\end{equation}
Using the fact $\sum\delta_i=0$ and $\delta=Max\{|\delta_i|\}$, we
obtain
\begin{equation}
e^{-\bar\mu} \le a_0\le e^{-\bar\mu}(1+\bar\mu^2\delta^2/2).
\end{equation}
Further, the fact that $\mu_-\le\bar\mu\le \mu_+$ leads to
\begin{equation}
e^{-\bar\mu_+} \le a_0\le e^{-\bar\mu_-}(1+\bar\mu^2\delta^2/2).
\end{equation}
This is the first inequality in Eq.(\ref{pbound0}). We have the
following equivalent form for $a_1=\frac{1}{N}\sum
\mu_ie^{-\mu_i}$:
\begin{equation}
a_1=\frac{1}{N}\bar\mu e^{-\bar\mu}\sum
(1+\delta_i)(1-\bar\mu\delta_i+\frac{1}{2}\bar\mu^2\delta_i^2-\cdots)
\end{equation}
This means
\begin{equation}
\bar\mu e^{-\bar\mu}(1-\bar\mu\delta^2)\le a_1\le\bar\mu
e^{-\bar\mu}
\end{equation}
which gives rise to
\begin{equation}
\mu_- e^{-\mu_-}(1-\mu_-\delta^2)\le a_1\le\mu_+ e^{-\mu_+},
\end{equation}
the second inequality of Eq.(\ref{pbound0}). Next we consider
$a_c=1-a_0-a_1=1-\frac{1}{N}\sum(e^{-\mu_i}+\mu_ie^{-\mu_i})$. The
As a result of Taylor expansion
\begin{equation}
a_1=1-e^{-\bar\mu}(1+\bar\mu-\frac{\delta^2\bar\mu^2}{2}+\cdots)
\end{equation}
which leads to
\begin{equation}
1-e^{-\bar\mu}-\bar\mu e^{-\bar\mu} \le a_c\le
1-e^{-\bar\mu}-\bar\mu
e^{-\bar\mu}+e^{-\bar\mu}\bar\mu^2\delta^2/2.
\end{equation}
Given the bounds of of $\bar\mu$, we have
\begin{equation}
1-e^{-\bar\mu_-}-\bar\mu_- e^{-\bar\mu_-} \le a_c\le
1-e^{-\bar\mu_+}-\bar\mu_+
e^{-\bar\mu_+}+e^{-\bar\mu_+}\bar\mu_+^2\delta^2/2.
\end{equation}
The derivations of the first two inequalities in
Eq.(\ref{pbound1}) are same with that of Eq.(\ref{pbound0}). We
only show the third one here. To obtain the lower bound, we have
\begin{equation}
\frac{\sum \mu_i'^2 e^{-\mu_i'}/N'}{\sum \mu_i^2 e^{-\mu_i}/N}\ge
\frac{\bar\mu'^2 e^{-\bar\mu'}}{(1+\delta^2)\bar\mu^2
e^{-\bar\mu}}.
\end{equation}
Therefore we have
\begin{equation}
a_c'\ge \frac{\bar \mu'^2
e^{\bar\mu-\bar\mu'}a_c}{\bar\mu^2(1+\delta^2)}\ge \frac{\mu_-'^2
(1-e^{-\bar\mu_-}-\bar\mu_-
e^{-\bar\mu_-})}{(1+\delta^2)\mu_+^2e^{\mu_-'-\mu_+}}.
\end{equation}
Given that $b_c'=\omega_c a_c'$, we arrive at the third inequality
of Eq.(\ref{pbound1}).

\begin{thebibliography}{99}
\bibitem{H03}
W.-Y.~Hwang, Phys. Rev. Lett. {\bf 91}, 057901 (2003).

\bibitem{Wang05}
X.-B.~Wang, Phys. Rev. Lett. {\bf 94}, 230503 (2005).

\bibitem{Wang05_2}
X.-B.~Wang, Phys. Rev. A {\bf 72}, 012322 (2005).

\bibitem{LMC05}
H.-K.~Lo, X.~Ma, and K.~Chen, Phys. Rev. Lett. {\bf 94}, 230504
(2005); X.~Ma {\em et al.}, Phys. Rev. A {\bf 72}, 012326 (2005).

\bibitem{HQph}
J.W.~Harrington {\em et al.}, quant-ph/0503002.
\bibitem{zei} R. Ursin et al, quant-ph/0607182.
\bibitem{scran}
V. Scarani, A. Acin, G. Robordy, N. Gisin, Phys. Rev. Lett. 92,
057901 (2004); C. Branciard, N. Gisin, B. Kraus, V. Scarani, Phys.
Rev. A 72, 032301 (2005).
\bibitem{kko} M. Koashi, Phys. Rev. Lett., 93, 120501(2004); K.
Tamaki, N. L\"ukenhaus, M. Loashi, J. Batuwantudawe,
quant-ph/0608082
\bibitem{BB84}C.H.~Bennett and
G.~Brassard, in {\em Proc.\ of IEEE Int.\ Conf.\ on Computers,
Systems, and Signal Processing (IEEE, New York, 1984)},
pp.~175-179.

\bibitem{GRTZ02}
N.~Gisin, G.~Ribordy, W.~Tittel, and H.~Zbinden,
Rev. Mod. Phys. {\bf 74}, 145 (2002).

\bibitem{DLH06}
M.~Dusek, N.~L\"utkenhaus, M.~Hendrych, "Quantum Cryptography", in
{\em Progress in Optics VVVX}, edited by E.~Wolf (Elsevier, 2006).


\bibitem{QKD}
M.~Bourennane {\em et al.}, F. Gibson, A. Karlsson, A. Hening,
P.Jonsson, T. Tsegaye, D. Ljunggren, and E. Sundberg, Opt. Express
{\bf 4}, 383 (1999); D.~Stucki {\em et al.}, D.~Stucki, N.~Gisin,
O.~Guinnard, G.~Ribordy and H.~Zbinden,  New. J. Physics, {\bf 4},
41, (2002); H.~Kosaka {\em et al.}, Electron. Lett. {\bf 39}, 1199
(2003); C.~Gobby, Z.L.~Yuan, and A.J.~Shields, Appl. Phys. Lett.
{\bf 84}, 3762 (2004); X.-F~Mo {\em et al.}, Opt. Lett. {\bf 30},
2632 (2005); G.Wu, J. Chen, Y. Li, L.-L. Xuand H.-P. Zeng,
quant-ph/0607099.
\bibitem{PNS}
B.~Huttner, N.~Imoto, N.~Gisin, and T.~Mor, Phys. Rev. A {\bf 51},
1863 (1995); H.P.~Yuen, Quantum Semiclassic. Opt. {\bf 8}, 939
(1996)
\bibitem{PNS1}G.~Brassard, N.~L\"utkenhaus, T.~Mor, and
B.C.~Sanders, Phys. Rev. Lett. {\bf 85}, 1330 (2000);
N.~L\"utkenhaus, Phys. Rev. A {\bf 61}, 052304 (2000);
N.~L\"utkenhaus and M.~Jahma, New J. Phys. {\bf 4}, 44 (2002).
\bibitem{GLLP04}
H.~Inamori, N.~L\"utkenhaus, D.~Mayers, quant-ph/0107017;
D.~Gottesman, H.K.~Lo, N.~L\"{u}tkenhaus, and J.~Preskill, Quantum
Inf. Comput. {\bf 4}, 325 (2004).

\bibitem{Lo06}
Y.~Zhao {\em et al.}, Phys. Rev. Lett. {\bf 96}, 070502 (2006);
Y.~Zhao {\em et al.}, quant-ph/0601168.

\bibitem{peng} C. Z. Peng et
al, quant-ph/0607129.

\bibitem{ron} D. Rosenberg et al, quant-ph/0607186.
\bibitem{137} X.-B. Wang et al, quant-ph/0609137.



\end{thebibliography}
 \end{document}